\documentclass[aps,prl,twocolumn,superscriptaddress]{revtex4}
\usepackage{epsf,graphicx}
\usepackage{amssymb}
\usepackage{amsmath}
\usepackage{latexsym,bm,array,amsfonts,multirow}
\usepackage{color}
\usepackage{ulem}
\begin{document}

\title{Decomposition of multilayer superconductivity with interlayer pairing}
\author{Yi-feng Yang}
\email[]{yifeng@iphy.ac.cn}
\affiliation{Beijing National Laboratory for Condensed Matter Physics and Institute of
Physics, Chinese Academy of Sciences, Beijing 100190, China}
\affiliation{University of Chinese Academy of Sciences, Beijing 100049, China}
\affiliation{Songshan Lake Materials Laboratory, Dongguan, Guangdong 523808, China}
\date{\today}

\begin{abstract}
We prove that multilayer superconductivity with interlayer pairing may naturally decompose into a series of weakly-coupled bilayer and trilayer superconducting blocks in order to minimize its total free energy. Our work is motivated by the recent proposal of interlayer pairing induced by the interlayer superexchange interaction of nearly half-filled $d_{z^2}$ orbitals in the bilayer and trilayer nickelate superconductors. We explore general properties of interlayer pairing superconductivity and perform systematic Ginzburg-Landau analyses of an effective multilayer model. For real materials, our results imply strong superconducting order parameter modulation and short coherence length along the $z$-axis (perpendicular to the layers). This reveals a unique feature of multilayer superconductivity with interlayer pairing and provides a basic framework for future experimental and theoretical investigations.
\end{abstract}
\maketitle

Layer structure has important influences on the properties of unconventional superconductors. In cuprates, $T_c$ reaches its record-high value in trilayer systems \cite{Scalapino2012a,Wang2023Science}, leading to the belief that multilayer may somehow promote the electron pairing. But in the recently-discovered multilayer nickelate superconductors, the maximum $T_c$ is reduced from about 80 K in the bilayer La$_3$Ni$_2$O$_7$ to 30 K in the trilayer La$_4$Ni$_3$O$_{10}$ \cite{Sun2023b,Hou2023,Zhang2023c,Li2024a,Zhu2023,Zhang2023m,Wang2024}. It has been shown that this opposite trend may be caused by their distinct pairing mechanisms, namely, intralayer pairing in cuprate superconductors and interlayer pairing in the bilayer and trilayer nickelate superconductors, owing to the different orbitals responsible for their pairing interactions \cite{Qin2024b}. While the cuprates have nearly half-filled $d_{x^2-y^2}$ orbitals with a dominant in-plane superexchange interaction \cite{LeTacon2011}, the bilayer and trilayer nickelates are governed mainly by the nearly half-filled $d_{z^2}$ orbitals with an interlayer superexchange interaction mediated by apical O \cite{Xie2024,Chen2024a}, which supports interlayer pairing for the superconductivity through hybridization with the nearly quarter-filled metallic $d_{x^2-y^2}$ bands \cite{Yang2023b,Qin2023b,Wang2024arxiv}.

Experimentally, the Ruddlesden-Popper (RP) phase of nickelates, La$_{n+1}$Ni$_{n}$O$_{3n+1}$, provides a material basis for studying the interlayer pairing superconductivity. Besides the bilayer and trilayer structures, other members ($n=4, 5, \infty$) have also been grown \cite{Li2020,Lei2017}, but superconductivity has not yet been reported, which possibly requires very high pressure. Other factors, such as the valence, oxygen vacancy \cite{Liu2023b,Dong2023}, and layer imbalance \cite{Luo2024,Tian2024}, might also have critical influence on the superconductivity. These are chemical properties that cannot be easily avoided and require tremendous efforts in material tuning. Nevertheless,  interlayer pairing superconductivity represents a future direction potentially different from intralayer pairing superconductivity such as the cuprates, and has rarely been explored. It is therefore intriguing to investigate general properties of multilayer superconductivity with interlayer pairing to provide some theoretical insight beforehand. 

Quite unexpectedly, we find that multilayer superconductivity with interlayer pairing has a natural tendency to decompose into bilayer and trilayer superconducting blocks separated by non-superconducting blocks. This leads to a strong order parameter modulation along the $z$ direction perpendicular to the layer plane. A small interlayer hopping may induce a weak Josephson coupling between these decoupled superconducting blocks, so that the whole structure may be viewed as a series of weakly-coupled bilayer and trilayer superconductors. Our observation provides a basic framework for future explorations of multilayer superconductivity with interlayer pairing.

\begin{figure}[t]
\centering
\includegraphics[width=0.7\linewidth]{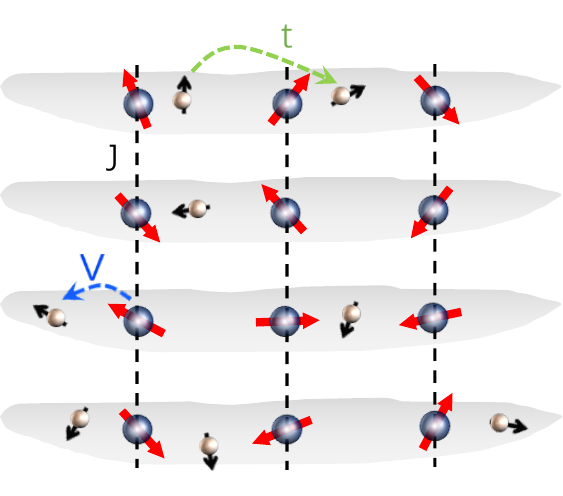}
\caption{Illustration of the $t$-$V$-$J$ model for multilayer superconductivity with interlayer pairing, where $t$ is the hopping parameter of the metallic band, $J$ is the interlayer superexchange interaction of the local pairing orbital, and $V$ is their hybridization.}
\label{fig1}
\end{figure}

We start with the following effective two-orbital multilayer $t$-$V$-$J$ model as illustrated in Fig. \ref{fig1} \cite{Yang2023b,Qin2023b,Wang2024arxiv,Qin2024b}:
\begin{eqnarray}
&H&=-\sum_{lijs}(t_{ij}+\mu\delta_{ij})c_{lis}^{\dagger}c_{ljs}-\sum_{lij}V_{ij}\left(c_{lis}^{\dagger}d_{ljs}+h.c\right)\notag\\
&&+J\sum_{ai}\bm{S}_{ai}\cdot\bm{S}_{a+1,i}-t_\perp\sum_{ais} \left(d_{ais}^\dagger d_{a+1,is}+h.c.\right),
\label{eq:H}
\end{eqnarray}
where $d_{lis}$ and $c_{lis}$ represent the local pairing orbital and the metallic orbital, respectively, $\bm{S}_{li}=\frac12\sum_{ss'}d_{lis}^{\dagger}\bm{\sigma}_{ss'}d_{lis'}$ is the spin density operator of the pairing orbital, $t_{ij}$ and $\mu$ are the in-plane hopping and the chemical potential of the metallic band, $V_{ij}$ is the renormalized in-plane hybridization between two orbitals, $t_\perp$ is the renormalized interlayer hopping of the pairing orbital, and $J$ is the interlayer superexchange interaction. We use $l=1,\cdots, L$ to denote the layers and $a=1,\cdots, L-1$. A constraint may be applied to exclude the double occupancy on the local orbital. The model is motivated by the bilayer and trilayer nickelates, but our results can be easily extended to general multilayer superconductors with interlayer pairing beyond this particular model. 

For clarity, we ignore other complications and only focus on the superconductivity. The superexchange term is decoupled as
\begin{equation}
J\bm{S}_{ai}\cdot\bm{S}_{a+1,i}\rightarrow \sqrt{2}\left(\bar{\Delta}^{(a)}_i\Phi_{i}^a+\bar{\Phi}^a_{i}\Delta^{(a)}_i\right)+\frac{8\bar{\Delta}^{(a)}_i\Delta^{(a)}_i}{3J},
\end{equation}
where $\Phi_{i}^a=\frac{1}{\sqrt{2}}\left(d_{ai\downarrow}d_{a+1,i\uparrow}-d_{ai\uparrow}d_{a+1,i\downarrow}\right)$ denotes the local interlayer singlet in the $a$-th block between the $a$-th and $(a+1)$-th layers and $\Delta_i^{(a)}$ is the corresponding pairing field. Ignoring the imaginary time dependence of the auxiliary fields, $\Delta_i^{(a)}(\tau)\rightarrow \Delta_i^{(a)}$, we obtain the action in the Nambu representation \cite{sm}:
\begin{equation}
S=\sum_{n}\bar{\psi}_n(-i\omega_n+O)\psi_n+\frac{8\beta}{3J}\sum_{ia}|\Delta_i^{(a)}|^2,\label{Stp}
\end{equation}
where $O$ is a matrix given by the model parameters and the auxiliary fields and
\begin{equation}
\bar{\psi}_n=\left(\bar{c}_{1\uparrow}, c_{2\downarrow}, \cdots, c_{1\downarrow}, \bar{c}_{2\uparrow}, \cdots, \bar{d}_{1\uparrow}, d_{2\downarrow}, \cdots, d_{1\downarrow}, \bar{d}_{2\uparrow}, \cdots \right),
\end{equation} 
with 
\begin{equation}
\begin{split}
\bar{c}_{ls}&=(\bar{c}_{l1s} (\tilde{s}i\omega_n),\cdots, \bar{c}_{lNs}(\tilde{s}i\omega_n)),\\
\bar{d}_{ls}&=(\bar{d}_{l1s} (\tilde{s}i\omega_n),\cdots, \bar{d}_{lNs}(\tilde{s}i\omega_n)).
\end{split}
\end{equation}
Here $\omega_n$ denotes the fermionic Matsubara frequency, $\tilde{s}=1$ ($-1$) for $s=\uparrow$ ($\downarrow$), and $N$ is the total number of lattice sites. Integrating out the fermionic degrees of freedom gives the effective action of the paring fields alone:
\begin{equation}
S_{\rm eff}(\{\Delta^{(a)}_i\})=\frac{8\beta}{3J}\sum_{ia}|\Delta_i^{(a)}|^2-\sum_n\text{Tr}\ln\left(-i\omega_n+O\right).
\end{equation}
The above formula may be further simplified, but numerical simulations are still too heavy for large $L$ and $N$. For simplicity, we perform the Ginzburg-Landau (GL) analysis for uniform static pairing fields: $\Delta^{(a)}_i\rightarrow \Delta^{(a)}$. The GL free energy density can be derived straightforwardly from the effective action, $f_\text{GL}=S_{\rm eff}/\beta N$, and takes the perturbative form \cite{Qin2024b}:
\begin{eqnarray}
f_\text{GL}&=&f_\text{GL}^{(0)}+ f_\text{GL}^{(2)}+O(t_\perp^4)\notag\\
&=&\sum_{a}\left[c_1|\Delta^{(a)}|^2+c_2|\Delta^{(a)}|^4+2c_2|\Delta^{(a)}|^2|\Delta^{(a+1)}|^2\right.\notag\\
&&\left.-h(\bar{\Delta}^{(a)}\Delta^{(a+1)}+c.c.)\right], \label{eq:fGL}
\end{eqnarray}
where $c_1$ and $c_2$ are temperature-dependent constant determined by the model parameters, and $h\propto t_\perp^2$ represents the interlayer Josephson coupling. The sum is over $a=1,\cdots, L-1$ and we have introduced artificially $|\Delta^{(0)}|^2=|\Delta^{(L)}|^2=0$ as the boundary conditions. The detailed forms of $c_i$ and $h$ depend on the model parameters and are given in the supplemental material \cite{sm}. By now, controversies still exist on the electronic structures and pairing mechanisms of the nickelate high-temperature superconductivity \cite{Yang2023b,Qin2023b,Wang2024arxiv,Nakata2017,Sakakibara2024PRB,DXYao2023,GMZhang2023,QHWang2023,JPHu2023,Zhang2023d,Lechermann2023,Christiansson2023,WWu2023,GSu2023,Heier2023,Kuroki2024,CJWu2024,ZYLu2024,WLi2024,DXYao2023tJ,FWang2024,TXiang2023,KJiang2024,WKu2024,YYCao2024}, but our conclusion is robust as far as only the interlayer pairing is concerned, although parameter tuning may destroy the superconductivity and invalidate our starting point.

To simplify the analyses, we first ignore $h$ and minimize $f_{\text{GL}}$ by $\partial f_\text{GL}/\partial \bar{\Delta}^{(a)}=0$. This gives $L-1$ equations for the superconducting order parameters $\Delta^{(a)}$:
\begin{equation}
\Delta^{(a)}\left(-c_0+|\Delta^{(a)}|^2+|\Delta^{(a-1)}|^2+|\Delta^{(a+1)}|^2\right)=0,
\label{GLsolution}
\end{equation}
where $c_0=-c_1/2c_2>0$ in the superconducting phase. It is not straightforward to write down immediately the GL solutions. For clarity, we discuss below how to obtain them step by step. Our conclusion is that multilayer superconductivity with interlayer pairing for $L\ge4$ will decompose into weakly-coupled bilayer and trilayer superconducting blocks. For even $L$, it decomposes into $L/2$ separated bilayer blocks, while for odd $L$, it decomposes into $(L-3)/2$ bilayer blocks and one trilayer block.

Our proof contains several steps.

(1) For the bilayer model $L=2$, there is only one block satisfying $\Delta^{(1)}\left(-c_0+|\Delta^{(1)}|^2\right)=0$, which has two solutions, $\Delta^{(1)}=0$ and $|\Delta^{(1)}|^2=c_0$. For $c_0>0$ ($c_1<0$), the latter solution has lower free energy density, $f_\text{GL}=-c_1^2/4c_2$, which gives a mean-field solution of the superconductivity with interlayer pairing in the bilayer nickelate.

(2) For the trilayer model $L=3$, there are two order parameters, $\Delta^{(1)}$ and $\Delta^{(2)}$, satisfying two GL equations:
\begin{equation}
\begin{split}
\Delta^{(1)}\left(-c_0+|\Delta^{(1)}|^2+|\Delta^{(2)}|^2\right)=0,\\
\Delta^{(2)}\left(-c_0+|\Delta^{(1)}|^2+|\Delta^{(2)}|^2\right)=0,
\end{split}
\end{equation}
which give four candidate solutions:
\begin{equation}
\begin{split}
\text{(a)}\ \ &\Delta^{(1)}=0,\ |\Delta^{(2)}|^2=c_0, \\
\text{(b)}\ \ &\Delta^{(2)}=0,\ |\Delta^{(1)}|^2=c_0, \\
\text{(c)}\ \ &|\Delta^{(1)}|^2+|\Delta^{(2)}|^2=c_0, \\
\text{(d)}\ \ &\Delta^{(1)}=\Delta^{(2)}=0,
\end{split}
\end{equation}
with the free energy density $f_\text{GL}=0$ for (d) and $f_\text{GL}=-c_1^2/4c_2$ for all three others. Among them, (a) and (b) are special cases of (c). They all have the same free energy density as the GL solution of the bilayer model. The uncertainty in (c) was first observed in our previous Monte Carlo simulations, revealing unexpected superconducting frustration between two blocks, which is a unique feature of the trilayer model. Introducing a small $h$ fixes the uncertainty to $|\Delta^{(1)}|^2=|\Delta^{(2)}|^2=c_0/2$ at sufficiently low temperature, whose reduced magnitude explains the reduction of $T_c$ in the trilayer nickelate \cite{Qin2024b}. Unlike cuprate superconductors where the pairing occurs within each CuO$_2$ layer and the $T_c$ is maximized in the trilayer structure, its reduction here reflects a fundamental distinction of the interlayer pairing superconductivity, where the two outer layers compete to form spin-singlet pairs with the same inner layer.

(3) For any finite $L\ge 4$, we first prove that its GL solution must contain non-superconducting blocks. If $\Delta^{(a)}\neq 0$ for $L-1$ blocks, Eq. (\ref{GLsolution}) would reduce to
\begin{equation}
-c_0+|\Delta^{(a)}|^2+|\Delta^{(a-1)}|^2+|\Delta^{(a+1)}|^2=0.
\label{solution2}
\end{equation}
Applying this to two neighboring blocks, we obtain immediately $|\Delta^{(a)}|^2=|\Delta^{(a+3m)}|^2$ for all $a$ and integer $m$ satisfying $0\le a+3m\le L$, so that the superconducting order parameters must repeat periodically every three blocks. The boundary conditions, $|\Delta^{(0)}|^2=|\Delta^{(L)}|^2=0$, then require $|\Delta^{(3m)}|^2=|\Delta^{(L-3m)}|^2=0$ as long as $0<3m<L$. For finite $L\ge4$, we have $m=1$ and $|\Delta^{(3)}|^2=|\Delta^{(L-3)}|^2=0$, which violates our assumption that all $\Delta^{(a)}$ are nonzero. We therefore conclude that the GL solution for any finite $L\ge 4$ must contain at least one non-superconducting block.

\begin{figure}[t]
\centering
\includegraphics[width=\linewidth]{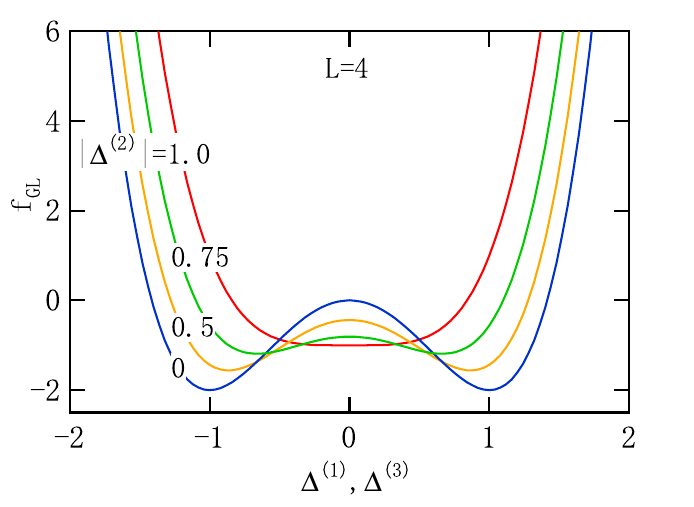}
\caption{Free energy density of the four-layer model with varying $\Delta^{(1)}=\Delta^{(3)}$ for different values of $|\Delta^{(2)}|$, showing the minima at $|\Delta^{(1)}|=|\Delta^{(3)}|=c_0$ on the $\Delta^{(2)}=0$ curve. The parameters are $h=0$, $c_1=-2$, and $c_2=1$, so that $c_0=-c_1/2c_2=1$ and the lowest free energy density $f_\text{GL}=-c_1^2/2c_2=-2$.}
\label{fig2}
\end{figure}

(4) Since each block is coupled only to its neighboring blocks in the free energy Eq. (\ref{eq:fGL}), a non-superconducting block splits the multilayer superconductivity into two decoupled  subsystems if only the superconductivity is concerned. Each subsystem as an independent multilayer model must also contain non-superconducting blocks if the number of its layers is greater than or equal to 4. This splits the subsystems until the whole system completely decomposes into a series of decoupled bilayer and trilayer superconducting blocks separated by non-superconducting blocks. As an example, we list all candidate solutions of the $L=4$ model:
\begin{equation}
\begin{split}
\text{(a)}\ \ &\Delta^{(2)}=0,\ |\Delta^{(1)}|^2=|\Delta^{(3)}|^2=c_0;\ f_\text{GL}=-c_1^2/2c_2,\\ 
\text{(b)}\ \ &\Delta^{(1)}=\Delta^{(2)}=0,\ |\Delta^{(3)}|^2=c_0;\ f_\text{GL}=-c_1^2/4c_2,\\
\text{(c)}\ \ &\Delta^{(1)}=\Delta^{(3)}=0,\ |\Delta^{(2)}|^2=c_0;\ f_\text{GL}=-c_1^2/4c_2,\\ 
\text{(d)}\ \ &\Delta^{(2)}=\Delta^{(3)}=0,\ |\Delta^{(1)}|^2=c_0;\ f_\text{GL}=-c_1^2/4c_2,\\ 
\text{(e)}\ \ &\Delta^{(1)}=0,\ |\Delta^{(2)}|^2+|\Delta^{(3)}|^2=c_0;\ f_\text{GL}=-c_1^2/4c_2,\\
\text{(f)}\ \ &\Delta^{(3)}=0,\ |\Delta^{(1)}|^2+|\Delta^{(2)}|^2=c_0;\ f_\text{GL}=-c_1^2/4c_2,\\
\text{(g)}\ \ &\Delta^{(1)}=\Delta^{(2)}=\Delta^{(3)}=0;\ f_\text{GL}=0.
\end{split}
\end{equation} 
All of them contain non-superconducting blocks. The solution (a) has the lowest free energy, which splits the $L=4$ structure into two superconducting blocks separated by a non-superconducting block. (b)(c)(d) reduce the model to a bilayer model, (e)(f) reduce it to a trilayer model, and (g) is a non-superconducting solution. Figure \ref{fig2} compares the free energy density as a function of $\Delta^{(1)}=\Delta^{(3)}$ for different values of $|\Delta^{(2)}|$. We see that $f_\text{GL}$ has the lowest value on the $\Delta^{(2)}=0$ curve rather than at a finite $\Delta^{(2)}$.

\begin{figure}[t]
\centering
\includegraphics[width=\linewidth]{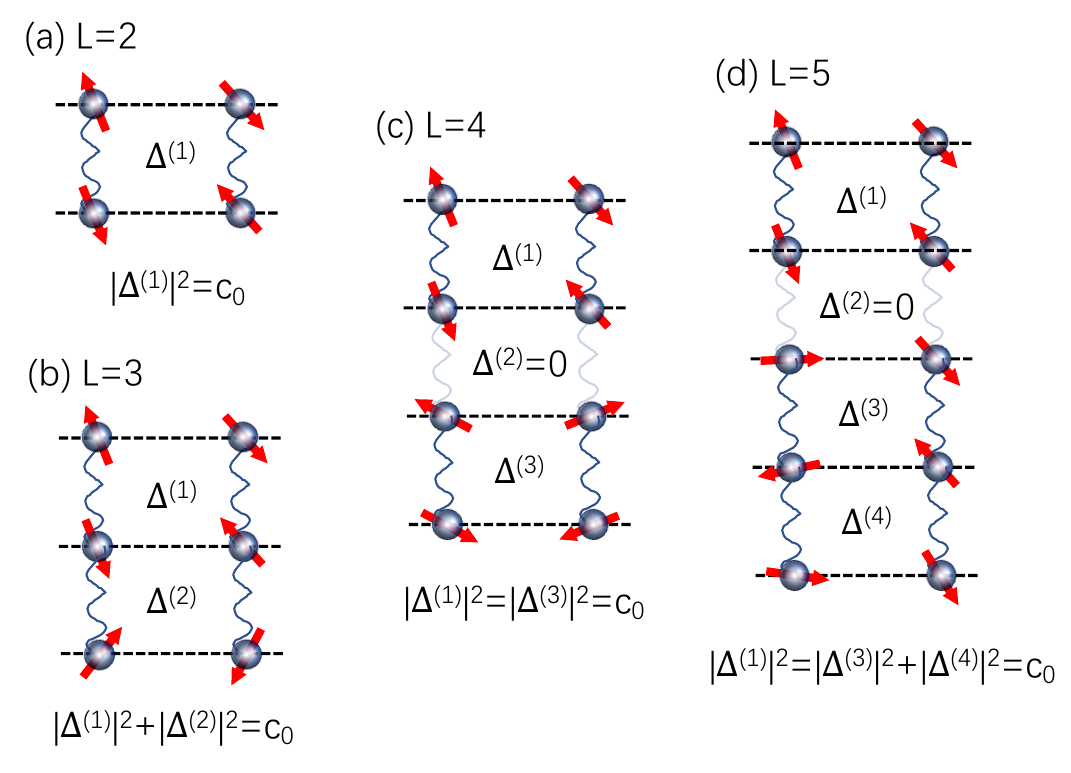}
\caption{GL decomposition of the multilayer superconductivity into a series of bilayer and trilayer superconducting blocks separated by non-superconducting blocks for $L=2$, 3, 4, 5. For $L=5$, there exists a second solution with $\Delta^{(3)}$=0 and $|\Delta^{(1)}|^2+|\Delta^{(2)}|^2=|\Delta^{(4)}|^2=c_0$.}
\label{fig3}
\end{figure}

(5) Among all candidate solutions, the GL decomposition should have the lowest free energy. But since the bilayer and trilayer superconducting blocks have the same free energy, minimizing the total free energy requires that the decomposition contains the largest number of decoupled superconducting blocks. For even $L$, this implies $L/2$ bilayer superconducting blocks, which gives the only GL solution with the lowest free energy. For odd $L$, this gives $(L-3)/2$ decoupled bilayer superconducting blocks plus one decoupled trilayer superconducting block. We have then $(L-1)/2$ equivalent solutions corresponding to different positions of the trilayer superconducting block. For an intuitive understanding, Fig. \ref{fig3} illustrates the decompositions for $L=2$, 3, 4, 5. 

(6) For infinite $L$, if all $\Delta^{(a)}$ are nonzero, the periodicity $|\Delta^{(a)}|^2=|\Delta^{(a+3m)}|^2$ allows one to simplify the free energy density to a three-block form:
\begin{equation}
\begin{split}
f_\text{GL}&=\sum_{a=1}^{3}\left(c_1|\Delta^{(a)}|^2+c_2|\Delta^{(a)}|^4+2c_2|\Delta^{(a)}|^2|\Delta^{(a+1)}|^2\right),\\
&=c_1\sum_{a=1}^{3}|\Delta^{(a)}|^2+c_2\left(\sum_{a=1}^{3}|\Delta^{(a)}|^2\right)^2,
\end{split}
\end{equation}
where $f_\text{GL}$ is the free energy density for three blocks and we have defined $|\Delta^{(4)}|^2=|\Delta^{(1)}|^2$. This gives a solution,
\begin{equation}
|\Delta^{(1)}|^2+|\Delta^{(2)}|^2+|\Delta^{(3)}|^2=c_0,
\end{equation}
with uncertainty as in the trilayer model. The free energy density (for three blocks) is then $f_\text{GL}=-c_1^2/4c_2$, the same as the bilayer (for one block) and trilayer (for two blocks) models. This decomposes the whole system into a series of three-block structure, whose free energy per block is higher than the bilayer decomposition. Thus, the infinite layer structure also favors a natural decomposition into a series of bilayer superconducting blocks.

(7) The above analyses assume $h=0$. Including interlayer hopping gives the Josephson coupling term $-h\bar{\Delta}^a\Delta^{(a+1)}$ in Eq. (\ref{eq:fGL}) \cite{Qin2024b}. For nearly half-filled $d_{z^2}$ orbitals, the interlayer hopping is strongly renormalized. Its magnitude is proportional to the hole density self-doped into the $d_{z^2}$ bonding orbital \cite{Yang2023b}. The parameter $h\propto t_\perp^2$ is therefore small and might only induce a weak proximity effect in the non-superconducting blocks and an even weaker coupling between neighboring superconducting blocks. As a result, the multilayer superconductivity is more like a series of weakly-coupled bilayer and trilayer superconducting junctions. If this is the case, it might host a high $T_c$ determined mainly by the bilayer superconducting blocks.

Taking together, we have proved that multilayer superconductivity with interlayer pairing in an ideal structure would intrinsically break down into weakly-coupled bilayer and trilayer superconducting blocks to minimize the total free energy. The bilayer and trilayer structures may therefore be regarded as the elementary blocks of interlayer pairing superconductivity. Note that the boundary conditions seem to play a key role for our conclusion. But this is actually not the case, as may be seen in the solution for infinite $L$, where we have discussed a three-block model with a periodic boundary condition. Ultimately, it is the number of decoupled superconducting blocks that plays the key role and should be maximized to give the lowest total free energy.

Our conclusion may hold for more general situations, although it is derived for an ideal structure with the same parameters for all layers. The free energy density Eq. (\ref{eq:fGL}) is a general consequence of the interlayering pairing term, $\bar{\Delta}^{(a)}_i\Phi_{i}^{a}+\bar{\Phi}^a_{i}\Delta^{(a)}_i$. For $t_\perp=0$, the $|\Delta^{(a)}|^2$, $|\Delta^{(a)}|^4$, and $|\Delta^{(a)}|^2|\Delta^{(a+1)}|^2$ terms represent all that can appear in the second and fourth-order perturbation expansion after integrating out the fermionic degrees of freedom. And the $h$ term represents the second-order contribution from the interlayer hopping. Thus, our observation reflects a general tendency of decomposition in multilayer superconductivity with interlayer pairing, at least on the mean-field perturbative level.

Real materials may also suffer from imbalance between layers or other material-specific factors, which may alter the parameters in the free energy and extend Eq. (\ref{eq:fGL}) to a more general form:
\begin{equation}
\begin{split}
f_\text{GL}&=\sum_{a=1}^{L-1}\left[c^{(a)}_1|\Delta^{(a)}|^2+c^{(a)}_2|\Delta^{(a)}|^4\right.\\
&\left. +2c^{(a)}_3|\Delta^{(a)}|^2|\Delta^{(a+1)}|^2-h^{(a)}(\bar{\Delta}^a\Delta^{(a+1)}+c.c.)\right],
\end{split}
\end{equation}
in which all parameters $c^{(a)}_i$ and $h^{(a)}$ are layer dependent. Again, we have defined $|\Delta^{(0)}|^2=|\Delta^{(L)}|^2=0$. For $h=0$, the GL solutions can be easily obtained numerically by minimizing the above free energy density. If the parameters are not changed significantly, the solutions should be close to those of the ideal model. Then, the superconductivity should still be intrinsically inhomogeneous and block dependent, which would reduce the superconducting coherence length along the $z$-axis.

Valence change may invalidate our effective model. In La$_{n+1}$Ni$_{n}$O$_{3n+1}$, the nominal valence of Ni ions is $\nu=3-1/n$, which gives $\nu=2.5$ for La$_3$Ni$_2$O$_7$, $2.67$ for La$_4$Ni$_3$O$_10$, $2.75$ for La$_5$Ni$_4$O$_{13}$, and $3$ for LaNiO$_3$. In La$_3$Ni$_2$O$_7$, the nearly half-filled $d_{z^2}$ orbitals and the nearly quarter-filled $d_{x^2-y^2}$ orbitals provide the pairing and metallic components, respectively. Their hybridization supports a two-component scenario for the high-temperature superconductivity \cite{Yang2023b,Qin2023b}. However, it remains to see if the variation of the Ni-valence with increasing $n$ might cause significant change in the property of the $d_{z^2}$ electrons and hence alter or even destroy the interlayer pairing. In any case, exploring possible high-temperature superconductivity through interlayer pairing of $d_{z^2}$ orbitals is still a feasible way to go beyond the cuprate scenario. Future experiments will overcome these challenging issues, grow more layered compounds with strong interlayer coupling, and tune them to achieve desired properties.

Last, we would like to emphasize again the peculiarity of our proposed minimal effective $t$-$V$-$J$ model \cite{Yang2023b,Qin2023b,Wang2024arxiv}. Different from the usual one-band model where the superexchange mechanism gives most probably a larger $J$ along the larger hopping direction \cite{Lin1997PRB}, the $t$-$V$-$J$ model separates the hopping and pairing terms into two hybridized orbitals, and thus allows for independent control of two key factors of the superconductivity. This lays the microscopic basis for  interlayer pairing and also points out a new route for exploring more high-temperature superconductors. More investigations may reveal even richer physics of the $t$-$V$-$J$ model \cite{Yang2024arxiv}.

To summarize, we have performed GL analyses of a multilayer model with interlayer pairing and proved on the mean-field perturbative level that its superconductivity may generally decompose into a series of weakly-coupled bilayer and trilayer superconducting blocks in order to minimize its total free energy. This implies intrinsic inhomogeneity and order parameter modulation along the $z$-axis. We hope more elaborate investigations will verify this unique feature of interlayer pairing superconductivity.

The author thanks Jiangfan Wang and Qiong Qin for useful discussions. This work was supported by the Strategic Priority Research Program of the Chinese Academy of Sciences (Grant No. XDB33010100), the National Natural Science Foundation of China (Grant No. 12174429), and the National Key Research and Development Program of China (Grant No. 2022YFA1402203).

\end{document}


\pagestyle{empty}

\title{Decomposition of multilayer superconductivity with interlayer pairing\\
	\vspace{0.2cm}
- Supplementary Material -}
\author{Yi-feng Yang}
\email[]{yifeng@iphy.ac.cn}
\affiliation{Beijing National Laboratory for Condensed Matter Physics and Institute of
Physics, Chinese Academy of Sciences, Beijing 100190, China}
\affiliation{University of Chinese Academy of Sciences, Beijing 100049, China}
\affiliation{Songshan Lake Materials Laboratory, Dongguan, Guangdong 523808, China}

\maketitle

We start from the following action for the original Hamiltonian:
\begin{equation}
\begin{split}
	S=&\int_{0}^{\beta}d\tau\left\{ \sum_{lijs}\bar{c}_{lis}(\tau)\left[(\partial_{\tau}-\mu)\delta_{ij}-t_{ij}\right]c_{ljs}(\tau)+\sum_{lis}\bar{d}_{lis}(\tau)\partial_{\tau}d_{lis}(\tau)\right.\\
	&\left.-\sum_{lijs}V_{ij}\left[\bar{d}_{lis}(\tau)c_{ljs}(\tau)+c.c. \right]-t_\perp\sum_{ais} \left[\bar{d}_{ais}(\tau) d_{a+1,is}(\tau)+c.c.\right]-\frac{3J}{4}\sum_{ai}\bar{\Phi}_{i}^{a}(\tau)\Phi_{i}^{a}(\tau)\right\}.
	\end{split}
\end{equation}
where $\Phi_{i}^a(\tau)=\frac{1}{\sqrt{2}}\left[d_{ai\downarrow}(\tau)d_{a+1,i\uparrow}(\tau)-d_{ai\uparrow}(\tau)d_{a+1,i\downarrow}(\tau)\right]$ is the interlayer pairing of local orbitals at site $i$ in the $a$-th block. The pairing term can be decoupled using the auxiliary pairing fields $\Delta_i^{(a)}(\tau)$:
\begin{equation}
-\frac{3J}{4}\bar{\Phi}_{i}^{a}(\tau)\Phi_{i}^{a}(\tau)\rightarrow \sqrt{2}\bar{\Delta}^{(a)}_i(\tau)\Phi_{i}^a(\tau)+\sqrt{2}\bar{\Phi}^a_{i}(\tau)\Delta^{(a)}_i(\tau)+\frac{8\bar{\Delta}^{(a)}_i(\tau)\Delta^{(a)}_i(\tau)}{3J},
\end{equation}
In the static approximation, $\Delta^{(a)}_i(\tau)\rightarrow \Delta^{(a)}_i$, this gives
\begin{equation}
S=\sum_{n}\bar{\psi}_n(-i\omega_n+O)\psi_n+\frac{8\beta}{3J}\sum_{ia}|\Delta_i^{(a)}|^2,\label{Stp}
\end{equation}
where 
\begin{equation}
\begin{split}
\bar{\psi}_n&=\left(\bar{c}_{1\uparrow}, c_{2\downarrow}, \bar{c}_{3\uparrow}, \cdots, c_{1\downarrow}, \bar{c}_{2\uparrow}, c_{3\downarrow}, \cdots, \bar{d}_{1\uparrow}, d_{2\downarrow}, \bar{d}_{3\uparrow}, \cdots, d_{1\downarrow}, \bar{d}_{2\uparrow}, d_{3\downarrow}, \cdots \right),\notag\\
\bar{c}_{ls}&=(\bar{c}_{l1s} (\tilde{s}i\omega_n),\cdots, \bar{c}_{lNs}(\tilde{s}i\omega_n)),\ \ \ \ \ \ \ \bar{d}_{ls}=(\bar{d}_{l1s} (\tilde{s}i\omega_n),\cdots, \bar{d}_{lNs}(\tilde{s}i\omega_n)).
\end{split}
\end{equation}
Here $\omega_n$ and the subscript $n$ denotes the fermionic Matsubara frequency and $\tilde{s}=1$ (-1) for $s=\uparrow$ ($\downarrow$). The matrix $O$ takes the form:
\begin{equation}
O=\left(\begin{array}{cccc}
A&0 &B&0\\
0&-A &0&-B\\
B&0 &M&D\\
0&-B &-D&M^*
\end{array}\right),
\end{equation}
with
\begin{equation}
\begin{split}
A=&\left(\begin{array}{ccccc}
-T&0&0&\cdots& 0\\
0&T&0&\cdots&0\\
0&0&-T&\cdots&0\\
\vdots &\vdots  & \vdots & \ddots&\vdots  \\
0&0&0&\cdots & (-1)^L T\\
\end{array}\right),~~~~B=\left(\begin{array}{ccccc}
-\mathcal{V}&0&0&\cdots& 0\\
0&\mathcal{V}&0& \cdots&0\\
0&0&-\mathcal{V}& \cdots&0\\
\vdots &\vdots  & \vdots & \ddots&\vdots  \\
0&0&0& \cdots & (-1)^L \mathcal{V}\\
\end{array}\right),\\
\\
D=&\left(\begin{array}{ccccc}
0&-T_\perp&0&\cdots& 0\\
T_\perp &0&T_\perp &\cdots& 0 \\
0 &-T_\perp  & 0 & \cdots & \vdots  \\
\vdots &\vdots  & \vdots & \ddots&(-1)^{L-1}T_\perp  \\
0&0&\cdots&(-1)^{L}T_\perp &0\\
\end{array}\right),~~M=\left(\begin{array}{ccccc}
0&M_1&0&\cdots&0\\
M_1^*&0&M_2^*&\cdots&0\\
0&M_2&0 & \cdots&\vdots\\
\vdots &\vdots  & \vdots & \ddots&\tilde{M}^*_{L-1}  \\
0&0&\cdots & \tilde{M}_{L-1}&0\\
\end{array}\right),
\end{split}
\end{equation}
where $T_{ij}=t_{ij}+\mu\delta_{ij}$, $(T_\perp)_{ij}=t_\perp\delta_{ij}$, $\mathcal{V}_{ij}=V_{ij}$, and $(M_a)_{ij}=\Delta_i^{(a)}\delta_{ij}$ with $i, j=1,\cdots, N$ and $a=1,\cdots,L-1$, and $\tilde{M}_{L-1}=M_{L-1}$ ($\tilde{M}_{L-1}=M^*_{L-1}$) if $L$ is odd (even).

Integrating out the fermionic degrees of freedom gives the GL free energy density:
\begin{equation}
f_\text{GL}(\{\Delta^{(a)}_i\})=S_{\rm eff}(\{\Delta^{(a)}_i\})/\beta N=\frac{8}{3JN}\sum_{ia}\bar{\Delta}_i^{(a)}\Delta_i^{(a)}-\frac{1}{\beta N}\sum_n\text{Tr}\ln\left(-i\omega_n+O\right).
\end{equation}
To simplify the derivation, we treat $t_\perp$ as a perturbation such that $f_\text{GL}=f_\text{GL}^{(0)}+ f_\text{GL}^{(2)}+O(t_\perp^4)$. For $t_{\perp}=0$, we have $D=0$ and
\begin{equation}
\text{Tr}\ln\left(-i\omega_n+O\right)=\text{Tr}\ln\left(-i\omega_n+O_1\right)+\text{Tr}\ln\left(-i\omega_n+O_2\right),
\label{st0}
\end{equation}
where 
\begin{equation}
O_1=\left(\begin{array}{cc}
A&B\\
B&M
\end{array}\right),~~~O_2=\left(\begin{array}{cc}
-A&-B\\
-B&M^*
\end{array}\right).
\end{equation}
Expanding over $\Delta_i^{(a)}$ (or $M$ and $M^*$) gives
\begin{equation}
\begin{split}
&\text{Tr}\ln\left(-i\omega_n+O_1\right)=	\text{Tr}\ln \left(\begin{array}{cc}
	-i\omega_n+A  & B \\
	B & -i\omega_n
\end{array}\right) -\frac{1}{2}\text{Tr}(P_1M)^2-\frac{1}{4}\text{Tr}(P_1M)^4+\cdots, \\
&\text{Tr}\ln\left(-i\omega_n+O_2\right)=	\text{Tr}\ln \left(\begin{array}{cc}
	-i\omega_n-A  & -B \\
	-B & -i\omega_n
\end{array}\right)-\frac{1}{2}\text{Tr}(P_2M^*)^2-\frac{1}{4}\text{Tr}(P_2M^*)^4+\cdots,
\label{exp}
\end{split}
\end{equation}
where
\begin{equation}
P_1=\left(\begin{array}{cccc}
		P_+ &0 &0&\cdots \\
		0 & P_- &0&\cdots \\
		\vdots  & \vdots & \ddots&\vdots  \\
		0 & 0 &\cdots& P_{-(-)^{L}}
		\end{array}\right),\quad P_2=\left(\begin{array}{cccc}
		P_- &0 &0&\cdots \\
		0 & P_+ &0&\cdots \\
		\vdots  & \vdots & \ddots&\vdots  \\
		0 & 0 &\cdots& P_{(-)^{L}} 
	\end{array}\right), \quad P_\pm=\frac{1}{-i\omega_n+\mathcal{V}(i\omega_n\pm T)^{-1} \mathcal{V}}.
\end{equation}
We have therefore
\begin{equation}
\begin{split}
	f_\text{GL}^{(0)}&=\frac{8}{3JN}\sum_{ia}|\Delta_{i}^a|^2+\frac{2}{\beta N}\sum_{na}\text{Tr}P_-M_a^*P_+M_a+\frac{1}{\beta N}\sum_{na}\text{Tr}
	(P_-M_a^*P_+M_a)^2\\
	&+\frac{1}{\beta N}\sum_{n}\sum_{a=1}^{L-2}\text{Tr}\left(P_-M_a^*P_+M_aP_-M_{a+1}^*P_+M_{a+1} +P_-M_a^*P_+M_{a+1}P_-M_{a+1}^*P_+M_a\right). \label{eq:FGL}
\end{split}
\end{equation}
Assuming uniform $(M_a)_{ii}=\Delta_i^{(a)}=\Delta^{(a)}$ yields
\begin{eqnarray}
	f_\text{GL}^{(0)}&=&c_1'\sum_a|\Delta^{(a)}|^2+c_2\left(\sum_{a}|\Delta^{(a)}|^4+2\sum_{a=1}^{L-2}|\Delta^{(a)}|^2|\Delta^{(a+1)}|^2\right),
\end{eqnarray}
where 
\begin{equation}
	c_1'=\frac{8}{3J}+\frac{2}{\beta N}\sum_n\text{Tr}(P_+P_-),~~~~~c_2=\frac{1}{\beta N}\sum_n\text{Tr}(P_+P_-P_+P_-).
\end{equation}
The second-order term $f_\text{GL}^{(2)}$ can be obtained by expanding the effective action (\ref{Stp}) to $O(t_\perp^2)$. We find
\begin{equation}
	f_\text{GL}^{(2)}=-\frac{2t_\perp^2}{\beta N}\sum_n\text{Tr}\left[\sum_a P_-^2 M_a^*P_+^2M_a+\sum_{a=1}^{L-2}\left(P_-^2 M_a^*P_+^2M_{a+1}+P_-^2 M_{a+1}^*P_+^2M_a\right)\right],
\end{equation} 
which, for uniform static order parameters $\Delta^{(a)}$, reduces to
\begin{equation}
	f_\text{GL}^{(2)}=-2t_\perp^2 c_3\left[\sum_a|\Delta^{(a)}|^2+\sum_{a=1}^{L-2}(\bar{\Delta}^a\Delta^{(a+1)}+c.c.)\right]
\end{equation}
with 
\begin{equation}
	c_3=\frac{1}{\beta N}\sum_n\text{Tr}(P_+P_+P_-P_-).
\end{equation}
The total free energy density is then
\begin{equation}
	f_\text{GL}=c_1\sum_a|\Delta^{(a)}|^2+c_2\left(\sum_{a}|\Delta^{(a)}|^4+2\sum_{a=1}^{L-2}|\Delta^{(a)}|^2|\Delta^{(a+1)}|^2\right)-h\sum_{a=1}^{L-2}(\bar{\Delta}^{(a)}\Delta^{(a+1)}+c.c.), \label{eq:fGL}
\end{equation}
where $c_1=c_1'-h$, and $h=2t_\perp^2c_3$ denotes the interlayer Josephson coupling. Introducing artificially $\Delta^{(0)}=\Delta^{(L)}=0$ gives the form presented in the main text.